\pdfoutput=1

\documentclass[a4paper]{jpconf}
\usepackage{graphicx}
\usepackage{amsmath}
\usepackage{booktabs}

\begin{document}
\title{Numerical Methods and the 4-point 2-loop Higgs amplitudes}

\author{S.~P.~Jones$^{a,}$\footnote{Speaker.}, B.~Ruijl$^b$}

\address{
$^a$Max Planck Institute for Physics, F\"ohringer Ring 6, 80805 M\"unchen, Germany \\
$^b$Institute for Theoretical Physics, ETH Z\"urich, 8093 Z\"urich, Switzerland}

\ead{sjones@mpp.mpg.de}

\begin{abstract}
Some of the difficulties faced when calculating multi-loop amplitudes with several mass scales are reviewed.
We then focus on one particular difficulty, the evaluation of the Feynman integrals, and introduce the program \texttt{pySecDec} which can be used to numerically compute such integrals.
Some of the new features and in particular the sector symmetry finder, which can help to reduce the number of sectors to be numerically integrated after sector decomposition, are described.
\end{abstract}

\section{Introduction}

In recent years there has been significant progress in the calculation of multi-loop amplitudes and related higher order quantities.
A prominent example is the computation of the N3LO QCD corrections to Higgs boson production in gluon-fusion~\cite{Anastasiou:2015ema}. However, processes with more legs but fewer loops, such as vector boson pair production at NNLO in QCD~\cite{Gehrmann:2015ora, vonManteuffel:2015msa, Caola:2015ila}, have recently also been analytically computed. Furthermore, progress has been made in the calculation of processes which depend on many scales (i.e. with internal massive particles or massive external legs), for example the planar integrals relevant for Higgs boson plus jet production~\cite{Bonciani:2016qxi} as well as several integrals relevant for mixed EW-QCD corrections and QED corrections including fermion masses~\cite{Bonciani:2016ypc,vonManteuffel:2017myy,DiVita:2017xlr,Mastrolia:2017pfy}. Advances have also been made in the computation of multi-loop processes with 5 external legs~\cite{Badger:2013gxa,Gehrmann:2015bfy,Papadopoulos:2015jft} and very recently even 6 and 7 external legs~\cite{Dunbar:2017nfy}.

Much of the recent progress has been driven by important insights into the analytical structure of Feynman integrals~\cite{Henn:2013pwa} as well as on-shell techniques, for example~\cite{Abreu:2017xsl}. However, progress has also been made in the context of predominantly numerical calculations as demonstrated by the numerical computation of the NLO QCD corrections to Higgs boson pair production~\cite{Borowka:2016ehy,Borowka:2016ypz}. An important open question is whether these numerical techniques can be used to compute further currently unknown processes such as the full Higgs boson plus jet amplitude or the Higgs plus Z-boson amplitude in gluon-fusion.

In these proceedings we briefly discuss some of the difficulties faced when computing multi-loop amplitudes with many mass scales. In Section~\ref{sec:2loop} we discuss the various stages of a multi-loop calculation carried out using mostly traditional techniques and highlight where the current technological bottlenecks are. In Section~\ref{sec:pysecdec} we focus on the particularly challenging issue of computing the relevant Feynman integrals and introduce a recent tool, \texttt{pySecDec}~\cite{Borowka:2017idc}, which is capable of computing such integrals numerically. Finally, in Section~\ref{sec:symmetry} we focus in some detail on one new feature of \texttt{pySecDec}, the sector symmetry finder, and describe its implementation.

\section{2-loop Amplitudes} \label{sec:2loop}

One standard method of computing multi-loop amplitudes consists of generating Feynman diagrams, inserting the Feynman rules (which produces integrals that must be evaluated), reducing the number of integrals using integration-by-parts identities~\cite{Chetyrkin:1981qh,Laporta:2001dd}, then evaluating the remaining so-called master integrals. This method was used for the calculation of Higgs boson pair production~\cite{Borowka:2016ehy,Borowka:2016ypz}. In more detail the steps taken (and compute time required) were:
\begin{enumerate}
\item decompose the amplitude into form factors and construct projectors (minutes),
\item generate Feynman diagrams (seconds),
\item apply the projectors and compute the amplitude (hours/days),
\item perform integral reduction (6+ months),
\item compute the master integrals (analytically: challenging, numerically: seconds/hours),
\item generate events and compute the differential cross-section (analytically: seconds?, numerically: hours/days).
\end{enumerate}
Roughly the same timings are valid for the calculation of Higgs boson plus jet production at NLO, though the final step is currently in progress and could in principle take significantly longer than estimated above. The calculation of the 2-loop gluon-fusion contribution to Higgs plus Z-boson production is yet more complicated due to the additional mass scale (Z-boson mass), this will impact the time taken to perform the integral reduction symbolically, though, fixing the top-quark and Higgs boson masses to numerical values may make the integral reduction more tractable. 

As can be seen from the timings above, by far the most time consuming step is the integral reduction. In the literature, several avenues are currently being explored to accelerate this step, see for example~\cite{vonManteuffel:2014ixa,Ita:2015tya,Larsen:2015ped,Maierhoefer:2017hyi}.
Another major bottleneck for the analytic approach is the computation of the master integrals.
The analytic evaluation of the master integrals is currently very challenging and so far only a subset of the integrals appearing in these processes is known~\cite{Bonciani:2015eua,Bonciani:2016qxi}. However, the numerical evaluation of the master integrals currently appears to be a practical alternative to the analytic approach and may be a promising research direction.

\section{Overview of \texttt{pySecDec}} \label{sec:pysecdec}

The program \texttt{pySecDec}~\cite{Borowka:2017idc} is designed to numerically compute dimensionally regulated parameter integrals.
It is a complete rewrite of its predecessor \texttt{SecDec} using only open source software and is intended to be as modular and extensible as possible. The initial symbolic/algebraic steps performed by \texttt{pySecDec} are carried out using \texttt{python}~\cite{python}, with extensive use of the packages \texttt{numpy}~\cite{numpy} and \texttt{sympy}~\cite{sympy}. Further symbolic manipulation and code generation are performed using \texttt{FORM}~\cite{Vermaseren:2000nd,Kuipers:2013pba,Ruijl:2017dtg}. The numerical integration of the resulting integrand functions is carried out in \texttt{c++11}~\cite{cpp11}. Extensive documentation for the code as well as tutorials and guides are provided using the Sphinx package~\cite{sphinx}.

The program \texttt{pySecDec} contains several new features in addition to many minor improvements, the key changes are:
\begin{itemize}
\item Support for an arbitrary number of regulators (not just $\epsilon$).
\item More flexible numerators (sums of contracted Lorentz vectors and inverse propagators).
\item Improvements to the handling of integrals without a Euclidean region.
\item Addition of a symmetry finder for detecting isomorphisms between sectors.
\item Use of \texttt{FORM}~\cite{Vermaseren:2000nd,Kuipers:2013pba,Ruijl:2017dtg} for code optimisation, this reduces the number of operations used to compute the integrand thus accelerating the numerical computation.
\item Automatic generation of a \texttt{c++} library which can be linked by external programs.
\end{itemize}

The input to \texttt{pySecDec} can be quite general parameter integrals. Much of what is discussed here and in Section~\ref{sec:symmetry} applies both to general parameter integrals and to the specific case of Feynman integrals. To rewrite a Feynman integral in terms of a dimensionally regulated parameter integral we first Feynman parametrise the integral. Concretely, an $L$-loop integral with $N$ propagators, $P_j$, raised to arbitrary powers, $\nu_j$, can be written as
\begin{align}
G 
& = \int_{- \infty}^{\infty} \left( \prod_{l=1}^L \frac{\mathrm{d}^D k_l }{i\pi^\frac{D}{2}} \right) \frac{1}{\prod_{j=1}^N P_j^{\nu_j} } \nonumber \\
& = (-1)^{N_\nu} \frac{\Gamma(N_\nu-LD/2)}{\prod_{j=1}^N \Gamma(\nu_j)} \int_0^\infty \left(\prod_{j=1}^N \mathrm{d}x_j \ x_j^{\nu_j-1} \right) \delta(1-\sum_{i=1}^N x_i)
\frac{\mathcal{U}^{N_\nu - (L+1)D/2}(x_1,\ldots,x_N)}{\mathcal{F}^{N_\nu-LD/2}(x_1,\ldots,x_N; s_1,\ldots,s_m)}. \label{eq:feynman}
\end{align}
Here we have introduced the Feynman parameters $x_1,\ldots,x_N$ and carried out the momentum integration. The symbols $s_1,\ldots,s_m$ represent invariants or masses and $N_\nu = \sum_{j=1}^N v_j$. The functions $\mathcal{U}$ and $\mathcal{F}$ are the 1st and 2nd Symanzik polynomials respectively. If inverse propagators are present, or if the numerator of the integral contains scalar products involving the loop-momenta, there may be an additional polynomial $\mathcal{N}$.

When calculating loop integrals the first step taken within \texttt{pySecDec} is to integrate out the Dirac delta.
For the decomposition strategies \texttt{iterative} and \texttt{geometric\_ku} this is done by introducing
\begin{equation}
\int_0^\infty d^N x = \sum_{l=1}^N \int_0^\infty d^N x \prod_{j=1, j \neq l}^N \theta(x_l \ge x_j),
\end{equation}
and performing one integration using the $\delta$-distribution. After this step we have $N$ integrals, which we refer to as primary sectors, each with one Feynman parameter set to one. 
For the decomposition strategy \texttt{geometric} the Cheng-Wu \cite{Cheng:1987ga,Smirnov:2006ry} theorem is instead used to integrate out the Dirac delta, this amounts to replacing the $\delta$-distribution in Eq.~\ref{eq:feynman} by $\delta(1-x_N)$.
After this so-called primary decomposition, the full sector decomposition algorithm is applied to each primary sector, for a description of the various strategies see Refs.~\cite{Heinrich:2008si,Borowka:2015mxa,Borowka:2017idc} and references therein. Finally, each sector is numerically integrated using one of the algorithms provided by CUBA~\cite{Hahn:2004fe} or, if the sector depends only on one variable, CQUAD~\cite{cquad,gsl}.

\section{Sector Symmetry Finder} \label{sec:symmetry}

In \texttt{pySecDec} a sector consists of one or more (potentially exponentiated) polynomials in the integration parameters. In the case of loop integrals, a sector consists of decomposed $\mathcal{U}$ and $\mathcal{F}$ polynomials, which depend on the Feynman parameters, raised to the appropriate powers according to Eq.~\ref{eq:feynman}, and, if a numerator is present, an additional $\mathcal{N}$ polynomial.

If sectors are equivalent up to permutations of the integration parameters then it is sufficient to compute one such sector and multiply the result by the number of equivalent sectors, this saves compute time during the numerical integration step. Since all integration parameters $x_i$ are integrated over the same domain $[0,1]$ we may freely switch the labels of the integration parameters within each sector without changing the result after integration. 
We have therefore implemented algorithms, described below, which can identify equivalent sectors up to such a relabelling. These algorithms work not with sectors but polynomials. To convert a sector into a single polynomial it suffices to multiply each polynomial within the sector by a unique label and sum the resulting polynomials.

To motivate the problem of identifying equivalent polynomials up to relabelling of variables consider the polynomials:
\begin{align}
\begin{split}
P_1 &=  x_1^2 x_2^1 x_3^1 + x_1^1 x_2^1 x_3^2, \\
P_2 &=  x_1^1 x_2^1 x_3^2 + x_1^1 x_2^2 x_3^1.
\end{split} \label{eq:polys}
\end{align}
In this case it is straightforward to see that the polynomials are equivalent up to relabelling of the variables $x_i$. By applying either permutation $S_1=(2\ 1)$ or $S_2=(2\ 3\ 1)$ to $P_2$ we obtain $P_1$. A brute force approach consists of applying all $n!$ permutations of the parameters $x_1, \ldots, x_n$ to one of the polynomials and then checking if the polynomials are equivalent. Such an algorithm has combinatorial complexity and becomes impractically slow for large problems.

In \texttt{pySecDec} there are four algorithms for identifying isomorphic sectors. Two of the algorithms, \texttt{iterative\_sort} and \texttt{light\_Pak\_sort}, are relatively quick to run but do not find all isomorphisms. The remaining algorithms \texttt{dreadnaut} and \texttt{Pak\_sort} find all identities but may take considerably longer to do so.

All algorithms presented in this section operate on the exponent list representation of multivariate polynomials. 
An exponent vector $\mathbf{e} = (e_1, e_2, \ldots, e_n) \in \mathrm{N}^n$ defines a monomial $\mathbf{x}^\mathbf{e} = x_1^{e_1} x_2^{e_2} \cdots x_n^{e_n}$. 
A term is the product of a non-zero coefficient $c$ and a monomial, i.e. $c \mathbf{x}^\mathbf{e}$.
The exponent list representation of a polynomial is a list of terms (coefficients and exponent vectors) appearing in the polynomial.  For example, the first polynomial in Eq.~\ref{eq:polys} has exponent list representation:
\begin{align}
P_1 &= x_1^2 x_2^1 x_3^1 + x_1^1 x_2^1 x_3^2 
\rightarrow
\bordermatrix {
                             & \mathrm{Coeff} \cr
\mathrm{Term\ 1} & 1 \cr
\mathrm{Term\ 2} & 1 \cr
}
\bordermatrix {
                           & x_1 & x_2 & x_3 \cr
 & 2    & 1     & 1   \cr
 & 1    & 1     & 2   \cr
} \label{eq:poly1}
\end{align}
In the exponent list representation simultaneously permuting the rows of the coefficient list and the exponent list corresponds to switching the order of terms in the polynomial, whilst permuting columns of the exponent list corresponds to switching the variable labels.

\subsection{Graph-based (\texttt{dreadnaut}) method} \label{sec:dreadnaut}

We begin by discussing the \texttt{dreadnaut} symmetry finder algorithm. The algorithm works by constructing a graph representation of the multivariate polynomial, bringing the graph into a canonical form by relabelling the vertices, hashing the canonical graph and then comparing the canonical graphs of polynomials with identical hashes. If two polynomials have identical canonical graphs they are equivalent up to permutations of the variable labels. The use of the \texttt{dreadnaut} symmetry finder is disabled by default in \texttt{pySecDec 1.2.2}, it can be enabled by setting \texttt{use\_dreadnaut = True} in the call to \texttt{make\_package} or \texttt{loop\_package}.

\begin{figure}
\begin{center}
\includegraphics[width=.45\textwidth]{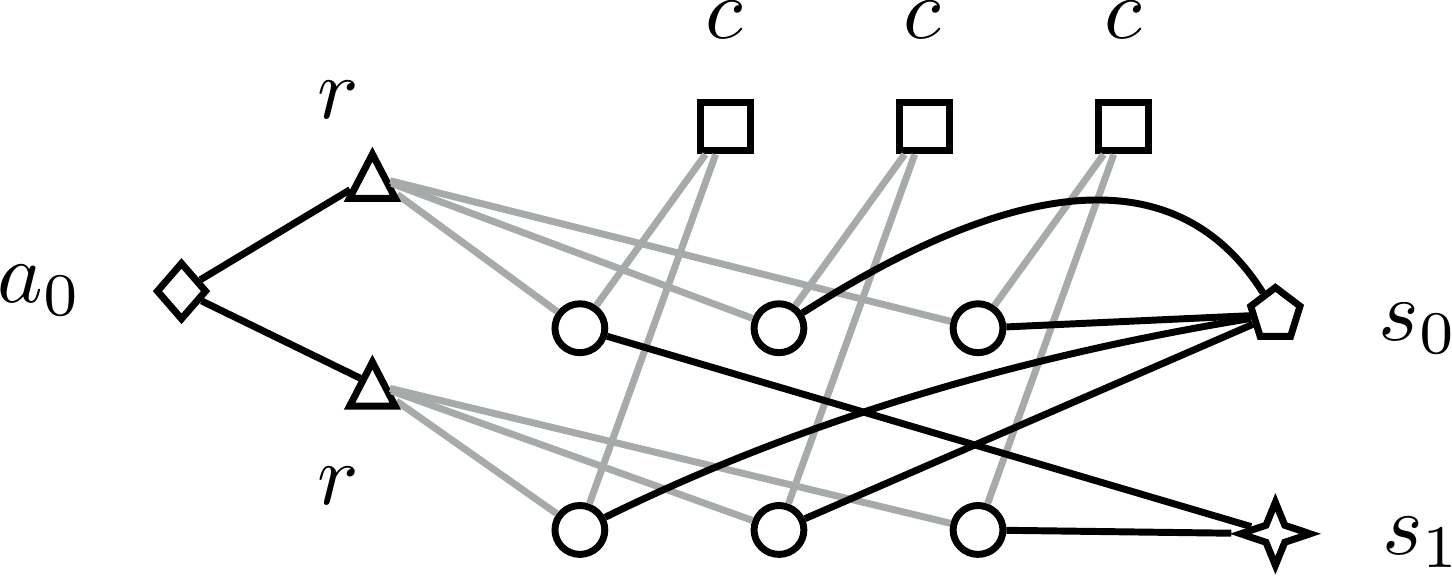}
\qquad
\fbox{\includegraphics[width=.45\textwidth]{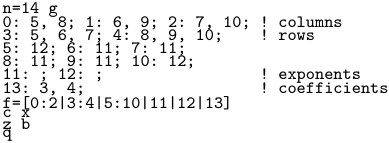}}
\end{center}
\caption{(Left panel) The graph corresponding to polynomial $P_1$ of Eq.~\ref{eq:poly1} constructed as described in Section~\ref{sec:dreadnaut}. (Right panel) The \texttt{dreadnaut} input corresponding to the graph.}
\label{fig:dreadnaut}
\end{figure}

In Figure~\ref{fig:dreadnaut} (left panel) we show the conversion of the polynomial $P_1$ in Eq.~\ref{eq:poly1} to a graph.
The algorithm we utilise for converting a polynomial into a graph is based on that used in Ref.~\cite{McKayPiperno} for identifying isotopy of matrices. The following steps are taken:
\begin{enumerate}
\item introduce a node of colour $c$ for each variable (column of the exponent matrix),
\item introduce a node of colour $r$ for each term (row of the exponent matrix),
\item introduce a node of colour $e$ for each entry in the exponent matrix, connect the entry node to the node of the corresponding column and to the node of the corresponding row,
\item for each unique exponent $0,\ldots,p$ appearing in the exponent matrix introduce a node of a new colour $s_0,\ldots,s_p$ and connect the exponent node to any entry node whose corresponding entry in the exponent matrix has this exponent,
\item for each unique coefficient $0,\ldots,m$ in the coefficient list introduce a node of a new colour $a_0,\ldots,a_m$ and connect it to the row node of any term which has this coefficient.
\end{enumerate}

To canonically label the graph we rely on the external program \texttt{dreadnaut}~\cite{McKayPiperno} which internally uses the graph algorithms of \texttt{nauty} and \texttt{Traces}. Usually the \texttt{dreadnaut} program can efficiently compute a canonical labelling of a graph, after which graphs that are isomorphic (the same except for vertex labels) become identical (exactly the same). Furthermore, after introducing a canonical labelling \texttt{dreadnaut} can output a graph hash (three 8-digit hex numbers). Isomorphic graphs always have the same hash, non-isomorphic graphs rarely have the same hash. These hashes can be used to quickly identify polynomials that may be identical. In Fig.~\ref{fig:dreadnaut} (right panel) we show the input to \texttt{dreadnaut} which performs these steps for the polynomial in Eq.~\ref{eq:poly1}; lines 1--8 define the graph, line 9 requests the canonical labelling, lines 10--11 instruct the program to output the graph hash and quit.

By default \texttt{dreadnaut} (and thus \texttt{pySecDec}) uses the McKay canonical graph labelling algorithm as implemented in \texttt{nauty}. Although this algorithm is known to have exponential complexity on some inputs \cite{complexity1, complexity2} it generally performs well and in practice we observe that it is typically much faster than a brute force algorithm.

\subsection{Pak algorithm}

An alternative method for identifying equivalent polynomials was described by Pak in Ref.~\cite{Pak:2011xt}. Instead of mapping the polynomial to a graph, the algorithm works by permuting the columns of the exponent matrix (which corresponds to relabelling variables) in order to maximize some metric. The permutations which maximize the metric can be considered to produce a ``canonical polynomial''. The equivalence of polynomials up to relabelling variables can be established by comparing their ``canonical'' representation. The Pak algorithm is enabled by default in \texttt{pySecDec 1.2.2}, it can be disabled by setting \texttt{use\_Pak = False} in the call to \texttt{make\_package} or \texttt{loop\_package}.
 
In principle this algorithm also has combinatorial complexity but beats a brute force algorithm by discarding more quickly obviously worse permutations.

\subsection{Performance}

\begin{figure}[h]
\begin{center}
\includegraphics[height=.12\textwidth]{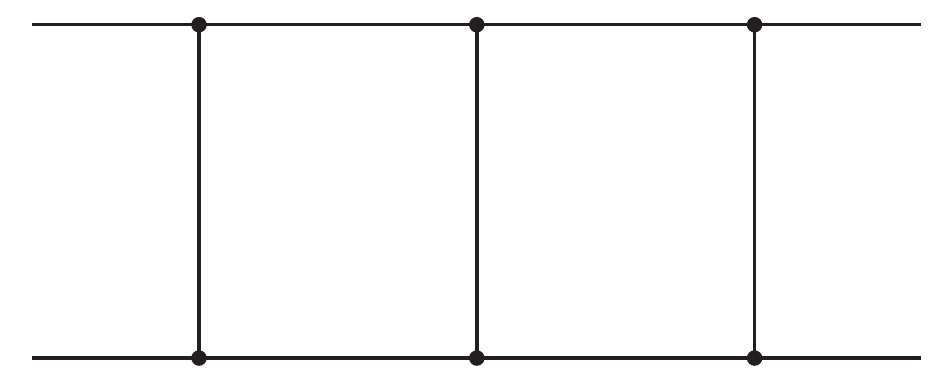}
\quad
\includegraphics[height=.12\textwidth]{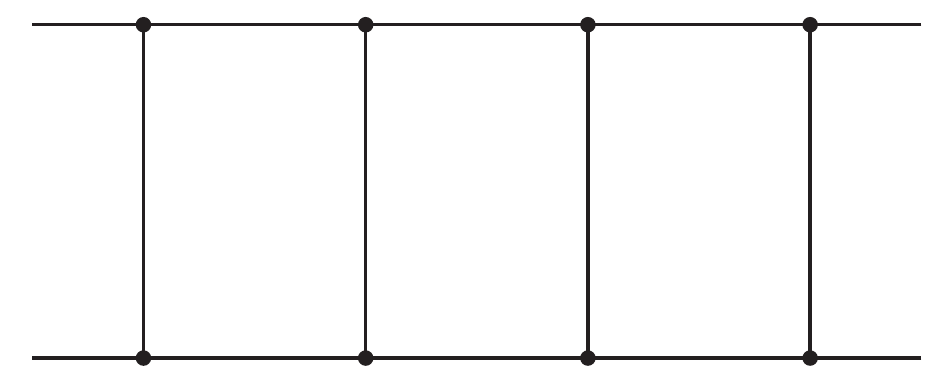}
\quad
\includegraphics[height=.12\textwidth]{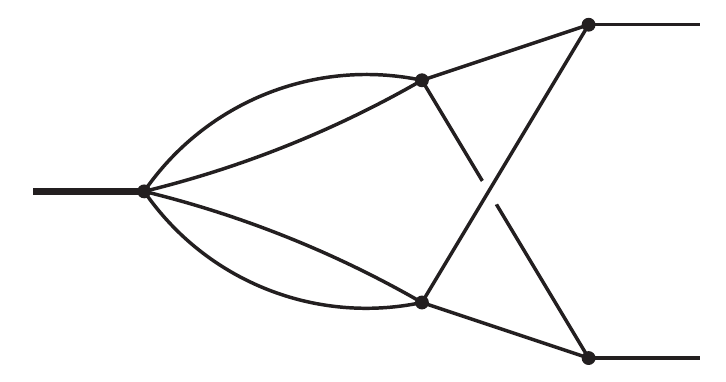}
\end{center}
\caption{Examples box2L, box3L and triangle4L which are used to benchmark the performance of the symmetry finder algorithms. All internal propagators are taken to be massless and massive external legs are displayed in bold.}
\label{fig:bench-diags}
\end{figure}

\begin{table}[]
\centering
\caption{The number of sectors before (total) and after (unique) applying the symmetry finder algorithms (\texttt{dreadnaut} or Pak) to each benchmark example. We show also the total wall time taken by each of the algorithms to identify all isomorphisms. All timings are taken using a single core of an Intel Core i7-7700 3.60 GHz CPU.}
\vspace{1em}
\label{tab:sectors}
\begin{tabular}{@{}lrrlrrlrr@{}}
\toprule
Example           & \multicolumn{2}{c}{Primary Sectors}                    & \phantom{abc}  & \multicolumn{2}{c}{Sectors}                            & \phantom{abc} & \multicolumn{2}{c}{Time {[}s{]}}     \\ 
           \cmidrule{2-3} \cmidrule{5-6} \cmidrule{8-9}
           & \multicolumn{1}{r}{Total} & \multicolumn{1}{r}{Unique} &  & \multicolumn{1}{r}{Total} & \multicolumn{1}{r}{Unique} &  & \multicolumn{1}{r}{\texttt{dreadnaut}}    & \multicolumn{1}{r}{Pak} \\ \midrule
triangle2L & 6                         & 4                          &  & 34                        & 21                         &  & 1.9                              & 0.1                   \\
triangle3L & 7                         & 3                          &  & 448                       & 124                        &  & 7.0                              & 0.54                    \\
triangle4L & 8                         & 3                          &  & 2848                      & 942                        &  & 50                               & 4.7                     \\
box2L      & 7                         & 3                          &  & 282                       & 110                        &  & 5.0                              & 0.42                    \\
box3L      & 10                        & 4                          &  & 9414                      & 3932                       &  & 770                              & 63                      \\ \bottomrule
\end{tabular}
\end{table}

In \texttt{pySecDec} the symmetry finding algorithms are applied first after primary sector decomposition and again after the complete sector decomposition. For benchmarking the symmetry finder we have selected 5 examples with between 2-4 loops and 3-4 legs. The examples triangle2L and triangle3L are taken from the \texttt{pySecDec} paper~\cite{Borowka:2017idc} and are described therein. In Figure~\ref{fig:bench-diags} we depict the examples box2L, box3L and triangle4L which have been introduced for the present benchmark.

The number of symmetries identified by the symmetry finder depends strongly on the problem. In particular, with no numerator present, the number of unique primary sectors can not be larger than the number of distinct graphs generated by pinching (removing) 1 propagator in each possible way. Many of the most complicated integrals appearing in Higgs boson pair production, Higgs boson plus jet production, and Higgs plus Z-boson production have few or no symmetries, though they typically have fewer sectors than the box2L example displayed here. For demonstration purposes, rather than selecting integrals from these processes, we have instead chosen integrals from the literature which have symmetries. 

In Table~\ref{tab:sectors}, for each example, we display the total number of primary sectors after primary decomposition, the number of unique primary sectors (remaining after symmetries have been identified), the total number of sectors obtained without identifying any symmetries, and the the number of unique sectors (after symmetry finding). For all examples we use the default \texttt{iterative} sector decomposition strategy. When counting the total number of sectors without identifying symmetries we also do not identify symmetries for primary sectors. Both the graph-based \texttt{dreadnaut} algorithm and the Pak algorithm identify exactly the same sector symmetries and so the total number of sectors after symmetry finding does not depend on the method used.

In practice, within \texttt{pySecDec}, we usually first apply the quick \texttt{iterative\_sort} algorithm followed by the \texttt{light\_Pak\_sort} algorithm then we apply a much slower full symmetry finder algorithm. In Table~\ref{tab:sectors} we present timings produced running only the Pak or \texttt{dreadnaut} algorithm without first applying the quicker but imperfect algorithms. We observe that for these examples the Pak algorithm is between 10-20 times faster than the \texttt{dreadnaut} algorithm.

\section{Conclusion}

We have outlined some of the key difficulties obstructing the straightforward computation of 4-point 2-loop Higgs amplitudes. The most time consuming step in the methods that we have currently applied to the computation of these amplitudes is the use of integration-by-parts identities to reduce the number of Feynman integrals appearing. Another key difficulty is the computation of the remaining integrals after this step. An overview of \texttt{pySecDec}, a program capable of computing such integrals numerically, was provided. The use of the recently introduced sector symmetry finder was described in detail. We have implemented two algorithms capable of identifying all sector symmetries, one is based on the canonical labelling of graphs and one, described previously by Pak, is based on the direct manipulation of the exponent list representation of multivariate polynomials. Although both algorithms identify precisely the same symmetries we find that the latter algorithm significantly outperforms the graph-based method for the examples considered here. The open source program \texttt{pySecDec} is publicly available and is provided with extensive documentation and usage guides.

\section*{Acknowledgements}
We would like to thank the \texttt{SecDec} Collaboration: Sophia Borowka, Gudrun Heinrich, Stephan Jahn, Matthias Kerner, Johannes Schlenk and Tom Zirke for many interesting discussions and the fruitful collaboration. SPJ is supported by the Research Executive Agency (REA) of the European Union under the Grant Agreement {PITN-GA}2012316704 (HiggsTools).

\end{document}